\documentclass[useAMS,usenatbib]{mn2e}

\usepackage{graphicx}
\usepackage{natbib}

\usepackage{amsmath}

\title[Mass flow rates in and outflow rates from AGN accretion discs]{Mass flow rates in and outflow rates from AGN accretion discs}
\author[M. Blank and W. J. Duschl]{Marvin Blank$^{1}$\thanks{E-mail:
mblank@astrophysik.uni-kiel.de (MB); wjd@astrophysik.uni-kiel.de (WJD)} and Wolfgang J.
Duschl$^{1,2}$\footnotemark[1]\\
$^{1}$Institut f\"{u}r Theoretische Physik und Astrophysik,
      Christian-Albrechts-Universit\"{a}t zu Kiel, Leibnizstr. 15, 24118 Kiel, Germany\\
$^{2}$Steward Observatory, The University of Arizona, 933 N. Cherry Ave., Tucson, AZ 85721, USA}

\begin{document}

\date{\today}

\pagerange{\pageref{firstpage}--\pageref{lastpage}} \pubyear{2012}

\maketitle

\label{firstpage}

\begin{abstract}
We derive analytical expressions for the mass flow rates in and from accretion
discs, taking into account the Eddington limit. This allows us to connect the
basic properties of outflows with those of the accretion flow. As an example,
we derive the radial surface density distribution in the accretion disc of Mrk~231.
\end{abstract}


\section{Introduction}\label{sec:intro}

It is now widely accepted that active galactic nuclei (AGN) are powered by accretion
of gas and dust via a disc on to the AGN's central black hole. 
This process releases sufficient energy to explain the observed AGN luminosities
\citep{1969Natur.223..690L, 1964SPhD....9..195Z}.
At high luminosities (radiation) pressure may become relevant and limit the accretion
rate to the so-called Eddington limit
\begin{equation}
 \dot{M}_{\rmn{Edd}} = \frac{M_{\rmn{bh}}}{\tau_{\rmn{S}}}
 \label{eq:edd}
\end{equation}
\citep{1921ZPhy....7..351E}, with the black hole mass $M_{\rmn{bh}}$, the Salpeter
time-scale $\tau_{\rmn{S}} = \eta \cdot 4.5 \cdot 10^{8}\,\rmn{yr}$ \citep{1964ApJ...140..796S}
and the accretion efficiency $\eta$, which is $\sim 10^{-1}$ for standard accretion discs
\citep{1973A&A....24..337S}.

Thus the black hole cannot swallow all the mass that is delivered by the accretion disc, 
what happens to the excess mass has been a subject of some intense debate.
\citet{2011MNRAS.413.1495d}, for instance, assume that this mass is driven from the AGN.
Such outflows were indeed observed from quasars (e.g., \citet{2005ASPC..331..361M} and \citet{2011MNRAS.410.1527H},
and, in particular for the context of this paper,  \citet{2011ApJ...729L..27R} from Mrk~231).

But some aspects of these outflows remain elusive, e.g., what amount of mass they contain
and what their power source is.
In this paper we present analytical and numerical investigations of mass flows in accretion
discs and mass outflows from accretion discs.
In section \ref{sec:model} we describe our accretion disc model, in section \ref{sec:supply-rates}
we derive an analytical description for the mass supply rate from accretion discs to black holes
which we verify in section \ref{sec:verify} by comparison with numerical calculations.
In section \ref{sec:outflows} we use this analytical description to calculate mass outflows and apply it,
as an example, to the mass outflow of Mrk~231 in section \ref{sec:outflows-mrk231}.

\section{Disc model}\label{sec:model}
For our accretion disc model we assume that
\begin{enumerate}
 \item the disc is rotationally symmetric,
 \item the disc is geometrically thin,
 \item the time derivative of the angular frequency is negligible, and
 \item the radial transport is dominated by gravitational and centrifugal forces.
\end{enumerate}

According to \citet{1981ARA&A..19..137P} the first three assumptions lead to one equation
that describes the time-dependent evolution of the accretion disc's surface density $\Sigma$.

\begin{equation}
 \frac{\partial \Sigma}{\partial t} + \frac{1}{s} \frac{\partial}{\partial s} \left[ \frac{\frac{\partial}{\partial s} \left( \nu \Sigma s^3 \frac{\partial \omega}{\partial s} \right)}{\frac{\partial}{\partial s} \left( s^2 \omega\right)} \right] = 0
 \label{eq:evolution}
\end{equation}

Here $\omega$ is the angular frequency, $\nu$ the viscosity of the gas and $s$ the distance to the central black hole,
respectively.

Using the fourth assumption and calculating the gravitational force at radius $s$ with the monopole approximation,
where just the disc mass within the radius $s$ and the black hole mass is considered and concentrated at
$s=0$ \citep[see, e.g.,][]{1997ApJ...480..167M}, allows us to determine the angular frequency
\begin{equation}
 \omega = \sqrt{\frac{G \left( M_{\rmn{bh}} + M_{\rmn{encl}}\left(s\right) \right) }{s^3}}
 \label{eq:omega}
\end{equation}
where
\begin{equation}
 M_{\rmn{encl}}(s) = 2 \pi \int\limits^s_{s_{\rmn{in}}} \Sigma(s') s' ds'
 \label{eq:enclmass}
\end{equation}
is the disc mass that is enclosed between the inner radius $s_{\rmn{in}}$ of the
accretion disc and the radius $s$.

As molecular viscosity is known to be too low to explain the observed quasar luminosities
\citep{1969Natur.223..690L}, turbulent viscosity is held responsible for powering quasars.
Due to the absence of a physical theory of turbulence several parametrizations have been
derived to describe the turbulent viscosity in accretion discs.
\citet{1973SvA....16..756S} and \citet{1973A&A....24..337S}
parametrize it in terms of what became known as the $\alpha$-viscosity: the product of the
disc's scale height $h$, the sound speed $c_{\rmn{s}}$ and a dimensionless parameter $\alpha$
which does not exceed unity. While this ansatz is very successful in describing non self-gravitating
discs, it runs into serious problems when self-gravity becomes relevant \citep{2000A&A...357.1123d}.
We follow them and use the so-called $\beta$-viscosity
\begin{equation}
 \nu = \beta s^2 \omega,
 \label{eq:viscosity}
\end{equation}
which contains the $\alpha$-viscosity in the
limiting case for non self-gravitating dissipation-limited discs.
The parameter $\beta$ is the inverse (critical) Reynolds number for the onset of turbulence of order $10^{-3}$ to $10^{-2}$.

Solving eq.~\ref{eq:evolution} allows us to calculate the mass supply rate $\dot{M}_{\rmn{d}}$
from the disc\footnote{The square brackets in eq.~\ref{eq:evolution} times $\left(-2 \pi\right)$ give the
     mass flow in the accretion disc, this evaluated at $s=s_{\rmn{in}}$ gives the mass supply rate of the
     disc. See, e.g., \citet{1981ARA&A..19..137P} for a detailed derivation.}.
Due to the Eddington limit, however, the black hole cannot necessarily swallow all
the mass that is delivered by the accretion disc.
Thus the accretion rate of the black hole, $\dot{M}_{\rmn{bh}}$, is limited to
\begin{equation}
 \dot{M}_{\rmn{bh}} = \min \left( \dot{M}_{\rmn{Edd}}, \dot{M}_{\rmn{d}} \right) \,.
 \label{eq:accrrate}
\end{equation}
The excess mass we assume to escape from the AGN region resulting in the mass outflow we are interested in.

\section{Mass supply rates of accretion discs}\label{sec:supply-rates}
In the following we will derive analytical expressions for the mass supply rate of accretion discs.
We assume an initial surface density profile 
\begin{equation}
 \Sigma_{0} \sim s^q
 \label{eq:sigmaprofile}
\end{equation}
with the exponent\footnote{For $q \leq -2$ the integral in eq.~\ref{eq:enclmass1} diverges, for $q \geq 1$ the viscous time-scale
     eq.~\ref{eq:timeII} is constant resp. decreasing with increasing radius $s$, which is unphysical.}
 $q \in (-2,+1)$ and find the disc mass enclosed within the radius $s$ according to eq.~\ref{eq:enclmass}:
\begin{equation}
 M_{\rmn{encl,0}}(s) = 2 \pi \int\limits_0^s \Sigma_{0}(s') s' ds' \sim s^{q+2}.
 \label{eq:enclmass1}
\end{equation}
Here we set $s_{\rmn{in}} = 0$ for simplicity.
Supposing that the disc's initial total mass $M_{\rmn{d,0}}$ is distributed between
$s=0$ and the disc's initial outer radius $s_{\rmn{out,0}}$ leads
\begin{equation}
  M_{\rmn{encl,0}}(s) = M_{\rmn{d,0}} \left( \frac{s}{s_{\rmn{out,0}}} \right)^{q+2} \, .
  \label{eq:enclmass0}
\end{equation}
As we neglect here the finite value of the inner radius $s_{\rmn{in}}$, this description is
applicable only for $s \gg s_{\rmn{in}}$.
But as the inner radius of accretion discs around black holes is of order of the Schwarzschild radius
$r_{\rmn{S}} \approx 10^{-13} \frac{M_{\rmn{bh}}}{\rmn{M}_{\odot}}\,\rmn{pc}$ and the extension of
accretion discs that concern us here is of order $10^{2\dots 3}\,$pc, this approximation does not
affect our results in any significant way.

We furthermore assume that the time $t$ the gas of the accretion disc needs to reach its centre
equals the viscous time-scale
\begin{equation}
  t = f \frac{s^2}{\nu} = \frac{f}{\beta \omega}
\end{equation}
with the dimensionless parameter $f$ of order unity.

In the following we will discern two cases for the ratio between the initial enclosed disc
mass and the initial central mass:
\begin{itemize}
\item In the \textit{low disc mass}\/ (LDM) case, the enclosed disc mass is small enough
      in comparison to the central object to be neglected for analytical approximations.
      This case applies \textit{(i)\/} if the total disc mass is small compared to the
      initial central mass right from the beginning of the simulation, or \textit{(ii)}\/ for
      sufficiently small radii, even if the enclosed disc mass for larger radii is non-negligible.
\item In the \textit{high disc mass}\/ (HDM) case, the enclosed mass is so high that, in
      comparison to the central object, the latter may be neglected for analytical approximations.
\end{itemize}
It is obvious that for any disc the innermost radial regions always belong the LDM case,
while at larger radii it depends on the mass distribution which case applies.

\subsection{The low disc mass case}

In the LDM case the angular frequency can be derived from eq.~\ref{eq:omega} by neglecting
the disc mass and taking into account the initial black hole mass $M_{\rmn{bh,0}}$ only:
\begin{equation}
   t = \frac{f_{\rmn{I}}}{\beta} \sqrt{ \frac{s^3}{G M_{\rmn{bh,0}}} } = \frac{f_{\rmn{I}}}{\beta \omega_{\rmn{I}}} \left( \frac{s}{s_{\rmn{out,0}}} \right)^{\frac{3}{2}}
  \label{eq:timeI}
\end{equation}
with
\begin{equation}
  \omega_{\rmn{I}} = \sqrt{\frac{G M_{\rmn{bh,0}}}{s^3_{\rmn{out,0}}}} \, .
\end{equation}
Rearranging and insertion of eq.~\ref{eq:timeI} into eq.~\ref{eq:enclmass0}
gives the total mass arriving at the accretion disc's centre in dependence of the time
\begin{equation}
 M_{\rmn{dI}}(t) = M_{\rmn{d,0}} \left(\frac{\beta \omega_{\rmn{I}}t}{f_{\rmn{I}}}\right)^{\frac{2}{3}\left(q+2\right)} \, ,
\end{equation}
whose derivative is the accretion disc's mass supply rate.
\begin{equation}
 \dot{M}_{\rmn{dI}}(t) = M_{\rmn{d,0}} \frac{2}{3} \left( q+2 \right) \left( \frac{\beta \omega_{\rmn{I}}}{f_{\rmn{I}}} \right)^{\frac{2}{3}\left(q+2\right)} t^{\frac{2q+1}{3}}
 \label{eq:massflow-anaI}
\end{equation}

\subsection{The high disc mass case}

In the HDM case the black hole mass in eq.~\ref{eq:omega} is neglected. With eq.~\ref{eq:enclmass0} for the enclosed disc mass,
the viscous time-scale is
\begin{equation}
 \begin{split}
   t = \frac{f_{\rmn{II}}}{\beta} \sqrt{ \frac{s^3}{G M_{\rmn{encl,0}}(s)} } = \frac{f_{\rmn{II}}}{\beta \omega_{\rmn{II}}} \left( \frac{s}{s_{\rmn{out,0}}} \right)^{\frac{1}{2}(1-q)}
 \end{split}
 \label{eq:timeII}
\end{equation}
with
\begin{equation}
  \omega_{\rmn{II}} = \sqrt{\frac{G M_{\rmn{d,0}}}{s^3_{\rmn{out,0}}}}.
\end{equation}
We derive the mass supply rate of the disc in the same manner than above:
\begin{equation}
 \dot{M}_{\rmn{dII}}(t) = M_{\rmn{d,0}} \frac{2 \left( q+2 \right)}{1-q} \left( \frac{\beta \omega_{\rmn{II}}}{f_{\rmn{II}}} \right)^{\frac{2 \left( q+2 \right)}{1-q}} t^{\frac{3 \left( q+1 \right)}{1-q}}.
 \label{eq:massflow-anaII}
\end{equation}

\subsection{The LDM/HDM transition}

Accretion from a massive disc on to the AGN's centre occurs in two phases:
During an initial phase I the accreted material comes only from the innermost
radii and thus the viscous time-scale for the inward movement is dominated by
the initial black hole mass, in which case the LDM approximation applies.
In the subsequent phase II, however,  it is dominated by the disc mass, including
the mass that has been accreted already. This is approximated by the HDM case.

Equating eqs.~\ref{eq:massflow-anaI} and \ref{eq:massflow-anaII} yields a transition time between the two phases:
\begin{equation}
 t_{\rmn{tr}} = \frac{f_{\rmn{tr}}}{\beta} \sqrt{\frac{s_{\rmn{out,0}}^3}{G M_{\rmn{tr}}}}
\end{equation}
with
\begin{equation}
 M_{\rmn{tr}} =  \left( M_{\rmn{bh,0}}^{q-1} M_{\rmn{d,0}}^{3} \right)^{\frac{1}{q+2}}
\end{equation}
and
\begin{equation}
 f_{\rmn{tr}} = \left[ \left( \frac{1-q}{3} \right)^{\frac{3}{2} \frac{1-q}{q+2}} f_{\rmn{I}}^{q-1} f_{\rmn{II}}^{3} \right]^{\frac{1}{q+2}} \, .
\end{equation}
$f_{\rmn{tr}}$ is of order unity as long as $q$ is not too small (see Fig.~\ref{fig:factors}).

\section{Verifying the mass supply rate formulas}\label{sec:verify}
Eq.~\ref{eq:evolution} describes the time- and space-dependent evolution of
the accretion disc's surface density $\Sigma$.
The solution of the set of eqs.~\ref{eq:edd} - \ref{eq:accrrate} allows us
to determine the mass supply rate $\dot{M}_{\rmn{d}}$ from the accretion disc,
which we will compare with our above analytical approach.
The initial condition for the surface density $\Sigma$ is given by eq.~\ref{eq:sigmaprofile},
leaving us with the parameters $\beta$ (viscosity), $q$ (initial surface density distribution),
$s_{\rmn{out,0}}$ (initial disc size), $M_{\rmn{d,0}}$ (initial disc mass), $M_{\rmn{bh,0}}$
(initial black hole mass) and $\eta$ (accretion efficiency).
We set $\eta = 10^{-1}$ throughout this paper leaving five parameters.

\begin{figure}
  \includegraphics[width=84mm]{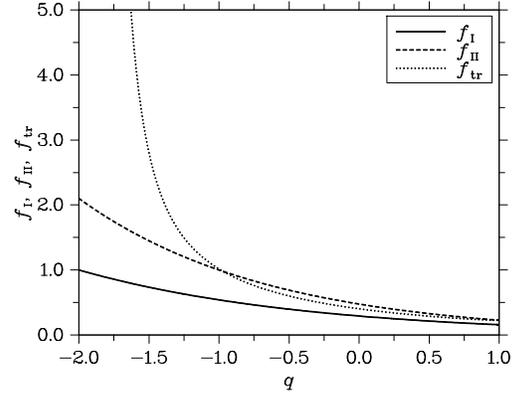}
  \caption{The factors $f_{\rmn{I}}$, $f_{\rmn{II}}$ and $f_{\rmn{tr}}$ as functions of $q$.}
  \label{fig:factors}
\end{figure}

\begin{table}
  \caption{Parameters of the reference model.}
  \label{tab:refmodel}
  \begin{tabular}{ll}
    \hline
    $\beta$           & $0.005$ \\
    $q$               & $-0.8$ \\
    $s_{\rmn{out,0}}$ & $1.5~\rmn{kpc}$ \\
    $M_{\rmn{d,0}}$    & $10^{10}~\rmn{M}_{\odot}$ \\
    $M_{\rmn{bh,0}}$   & $10^{5}~\rmn{M}_{\odot}$ \\
    \hline
  \end{tabular}
\end{table}

We carry out the time integration by applying an implicit Crank-Nicolson finite
differences scheme \citep{1947PCPS...43...50C}.
The radial calculation domain extends from an inner radius $s_{\rmn{in}}$, which
we set to $10^{-4}\,\rmn{pc}$ for all calculations presented in this paper. 
Tests showed that moderate variations of this value have almost no effect on the results.
Furthermore, we follow \citet{2011MNRAS.413.1495d} and set the outer radius of the
\textit{computational}\/ domain to $s^2_{\rmn{out,0}}\rmn{/}s_{\rmn{in}}$.
This is much larger than the initial \textit{physical}\/ outer radius of
the accretion disc and minimizes the influence of the outer boundary condition
on the temporal evolution of the model calculations.
For numerical reasons, but with negligible influence on the numerical results,
we initially distribute $10^{-3}$ of the initial disc mass between the disc's
outer radius and the numerical outer radius.
Between the inner and the outer numerical radius 150 grid points are distributed logarithmically.
Boundary conditions are applied by setting the surface density to zero at
the inner numerical radius which is equivalent to a vanishing torque,
and by setting the mass flow to zero at the outer numerical radius.

\begin{figure*}
  \includegraphics[width=168mm]{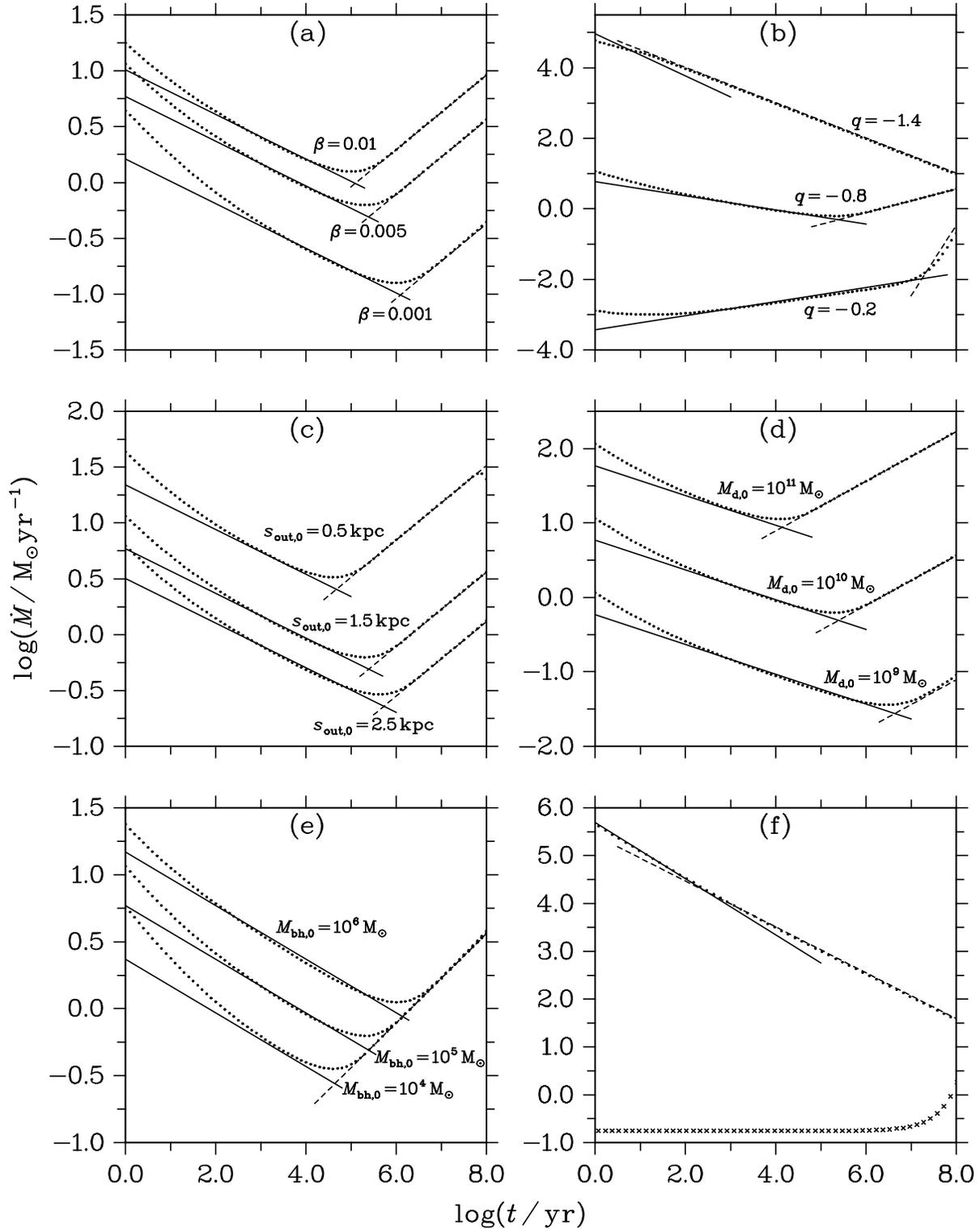}
  \caption{Mass supply rates of accretion discs and Eddington limit as functions of time. The dotted lines are
           the mass supply rates of the numerical calculations, the solid lines correspond to eq.~\ref{eq:massflow-anaI}
           and the dashed lines to eq.~\ref{eq:massflow-anaII}.
           The crossed line is the Eddington limit (Fig.\,(f) only).
           In Fig.\,(a) we varied the $\beta$-parameter, in Fig.\,(b) the exponent $q$, in Fig.\,(c)
           the disc's initial outer radius $s_{\rmn{out,0}}$, in Fig.\,(d) the initial disc mass $M_{\rmn{d,0}}$
           and in Fig.\,(e) the initial black hole mass $M_{\rmn{bh,0}}$ from the reference model, respectively.
           Fig.\,(f) shows numerical and analytical mass supply rates and the Eddington limit for our model parameters of Mrk~231.}
  \label{fig:massflows}
\end{figure*}

We define a reference model with the parameters given in Table~\ref{tab:refmodel},
in Figs.~\ref{fig:massflows}\,(a)-(e) we vary one of the parameters
$\beta$, $q$, $s_{\rmn{out,0}}$, $M_{\rmn{d,0}}$ and $M_{\rmn{bh,0}}$ from the ones of the reference model, respectively.
We found the mass supply rate formulas \ref{eq:massflow-anaI} and \ref{eq:massflow-anaII} fit well
with the numerical results for 
\begin{equation}
 f_{\rmn{I}} = a_{\rmn{I}}^{q+2}
\end{equation}
and
\begin{equation}
 f_{\rmn{II}} = a_{\rmn{II}}^{-\left(q+1\right)}
\end{equation}
with $a_{\rmn{I}}=0.54$ and $a_{\rmn{II}}=2.1$.
As these factors are of order unity (see Fig.~\ref{fig:factors}), however, they have
no crucial effect on our results.

The deviations for early times in Figs.~\ref{fig:massflows} (up to $10^{2}\,\rmn{yr}$)
are due to setting the accretion disc's inner radius to zero in the analytical description.
For large times (after $10^{8}\,\text{yr}$) deviations are expected because, in contrast to
the numerical results,  eq.~\ref{eq:enclmass0} does not take into account
that the disc has a finite radius at $s_{\rmn{out,0}}$.

\section{Mass outflows}\label{sec:outflows}
For the following order-of-magnitude calculations we use an average mass supply rate for phase II,
rather than the one derived in eq.\ \ref{eq:massflow-anaII}:
\begin{equation}
 \dot{M}_{\rmn{dII}}^{\rmn{avg}} = \frac{M_{\rmn{dII}}(t)}{t} = M_{\rmn{d,0}} \left( \frac{\beta \omega_{\rmn{II}}}{f_{\rmn{II}}} \right)^{\frac{2 \left( q+2 \right)}{1-q}} t^{\frac{3 \left( q+1 \right)}{1-q}} \, .
 \label{eq:avg}
\end{equation}
The average rate for phase I may be calculated in a similar manner.

As we will show in section \ref{sec:outflows-mrk231}, for not too large times and for disc masses much
larger than initial black hole masses the mass supply rate of the disc by far exceeds the Eddington limit
\begin{equation}
 \dot{M}_{\rmn{d}} \gg \dot{M}_{\rmn{Edd}} \, ,
 \label{eq:comparation}
\end{equation}
so almost all mass delivered by the disc escapes from the AGN and the excess mass flow, i.e.,
the difference between what is delivered by the disc and the Eddington limit, can be approximated as 
\begin{equation}
 \dot{M}_{\rmn{exc}} \approx M_{\rmn{d,0}} \left( \frac{\beta \omega_{\rmn{II}}}{f_{\rmn{II}}} \right)^{\frac{2 \left( q+2 \right)}{1-q}} t^{\frac{3 \left( q+1 \right)}{1-q}} \, .
 \label{eq:excess}
\end{equation}
This mass we assume to escape from the system as an outflow,
the lifetime of which can be estimated as
\begin{equation}
 t_{\rmn{lt}} = \frac{R}{v}
\end{equation}
with $R$ the extension and $v$ the velocity of the outflow, respectively. Thus eq.~\ref{eq:excess} gives
\begin{equation}
 \dot{M}_{\rmn{exc}} \left( \frac{v}{R} \right)^{\frac{3 \left( q+1 \right)}{1-q}} = M_{\rmn{d,0}} \left( \frac{\beta \omega_{\rmn{II}}}{f_{\rmn{II}}} \right)^{\frac{2 \left( q+2 \right)}{1-q}}
 \label{eq:avg2}
\end{equation}
which connects the parameters of the outflow, $\dot{M}_{\rmn{exc}}$, $R$ and $v$,
with the parameters of the accretion disc, $\beta$, $q$, $s_{\rmn{out,0}}$ and $M_{\rmn{d,0}}$.

$\dot{M}_{\rmn{exc}}$ can be determined by measuring the velocity, the extension,
the solid angle and the column density of the outflow.
For different approaches to determine $\dot{M}_{\rmn{exc}}$ from these quantities, see, for instance,
\citet{2010ApJ...724.1430S} (thin-shell model) and \citet{2005ApJS..160..115R} (thin- and thick-shell model).

\section{Mass outflow from Mrk~231}\label{sec:outflows-mrk231}
Mrk~231 is classified as ultra-luminous infrared galaxy \citep{1972ApJ...176L..95R}
and Seyfert~I galaxy \citep{1973ApJ...183...29W}.
It hosts an accretion disc with a mass of $3.15 \cdot 10^{10}\,\rmn{M}_{\odot}$,
an outer radius of $1.7\,\rmn{kpc}$ \citep{1998ApJ...507..615D} and a central unresolved
mass of $\la 8 \cdot 10^{6}\,\rmn{M}_{\odot}$ \citep{2005MNRAS.364..353R}, which we assume
to be the central black hole's mass.
For the $\beta$-parameter we adopt the value $\beta=0.005$ from our reference model.

\citet{2011ApJ...729L..27R} recently described a massive, wide-angle outflow from Mrk~231
and estimate an extension of the outflow of order $\sim 3\,\rmn{kpc}$, a velocity of $\sim 10^3\,\rmn{km}\,\rmn{s}^{-1}$
and a mass outflow rate\footnote{More precisely, \citet{2011ApJ...729L..27R} estimate the mass outflow rate as
     $\dot{M} = 420 \, (R/3~\rmn{kpc}) \, [0.1/(N(\rmn{Na~I})/N(\rmn{Na}))] \,\rmn{M}_{\odot}\,\rmn{yr}^{-1}$ where $N(\rmn{Na~I})$ and
     $N(\rmn{Na})$ are the column densities of Na~I and Na, respectively, and $R$ is the extension of the outflow.
     We adopt their suggested values of $N(\rmn{Na~I})/N(\rmn{Na})=0.1$ and $R = 3~\rmn{kpc}$.}
$420\,\rmn{M}_{\odot}\,\rmn{yr}^{-1}$.

With these data we write eq.~\ref{eq:avg2} as
\begin{equation}
 \begin{split}
   \left( \frac{\dot{M}_{\rmn{exc}}}{420~\rmn{M}_{\odot}~\rmn{yr}^{-1}} \right)^{1-q} \biggl( \frac{v}{1000~\rmn{km}~\rmn{s}^{-1}} \biggr)^{3\left(q+1\right)} \left( \frac{R}{3~\rmn{kpc}} \right)^{-3\left(q+1\right)} \\
   = g(q) \left( \frac{\beta}{0.005} \right)^{2\left(q+2\right)} \left( \frac{M_{\rmn{d,0}}}{3.15\cdot10^{10}~\rmn{M}_{\odot}} \right)^{3} \left( \frac{s_{\rmn{out,0}}}{1.7~\rmn{kpc}} \right)^{-3\left(q+2\right)}
 \end{split}
 \label{eq:detq}
\end{equation}
with
\begin{equation}
 g(q) = 9.85 \cdot 2.43^q \cdot 10^{-10-7q} \cdot f_{\rmn{II}}^{-2q-4} \, .
\end{equation}

For Mrk~231 the brackets in eq.~\ref{eq:detq} equal unity and by solving the equation $g(q)=1$ we can determine
$q\approx-1.38$ and thus the (time-averaged) surface density profile in Mrk~231 as $\Sigma \propto s^{-1.38}$.

The corresponding mass flows (eqs.~\ref{eq:massflow-anaI} and \ref{eq:massflow-anaII}) we plot in Fig.~\ref{fig:massflows}\,(f)
and compare them with numerical solutions of this model to show the agreement of numerical and analytical description
also for these parameters.
Additionally Fig.~\ref{fig:massflows}\,(f) shows that the Eddington limit is indeed much smaller than the
disc's mass supply rate for not too large times as we assumed in the previous section.

The transition time $t_{\rmn{tr}}\approx300\,\rmn{yr}$ is small compared to the lifetime 
of the outflow of $t_{\rmn{lt}}\approx3\cdot10^{6}\,\rmn{yr}$,
which justifies the use of the phase~II model for describing the mass outflow from Mrk~231.\footnote{Using the
phase~I model gives $t_{\rmn{tr}}\approx10\,\rmn{yr}$ and likewise suggests the sole use of the phase~II model.}

\section{Summary}
We derived analytical expressions for the mass supply rate of accretion discs
which depend on basic disc parameters, black hole mass and time
and verified them by comparison with the results of numerical calculations.

When the Eddington limit is much lower than the mass supply rate of the accretion disc,
this allows us to constrain the properties of mass outflows.

With these analytical expressions we are able to connect the basic
properties of the outflow ($\dot{M}_{\rmn{exc}}$, $R$ and $v$)
and of the accretion disc ($\beta$, $q$, $s_{\rmn{out,0}}$, $M_{\rmn{d,0}}$).
This relationship can be used to estimate basic properties of quasars.
As an example, we applied it to Mrk~231 and found the surface density
distribution of its accretion disc to be $\propto s^{-1.38}$.

As in our scenario the outflow is driven by the limitation of the black hole's accretion
rate through the Eddington limit, we conclude that, at least in such cases,
(radiation) pressure is capable of driving huge quasar mass outflows.

\bibliographystyle{mn2e}
\bibliography{manuscript}

\bsp

\label{lastpage}

\end{document}